\documentclass[twocolumn,english,floatfix,aps,prl]{revtex4}
\usepackage[T1]{fontenc}
\usepackage[latin1]{inputenc}
\usepackage[usenames]{color}

\makeatletter

\usepackage{babel}
\usepackage{graphicx}
\makeatother
\begin{document}

\preprint{This line only printed with preprint option}

\title{Characteristic BEC scaling close to Quantum Critical Point in BaCuSi$_2$O$_6$}

\author{S. E. Sebastian$^1$, P. A. Sharma$^2$, M. Jaime$^2$, N. Harrison$^2$, V. Correa$^2$, L. Balicas$^3$, N. Kawashima$^4$, C. D. Batista$^5$, I. R. Fisher$^1$}

\affiliation{$^1$Geballe Laboratory for Advanced Materials and
Department of Applied Physics, Stanford University, Stanford, CA
94305}

\affiliation{$^2$MST-NHMFL, Los Alamos National Laboratory, Los
Alamos, NM 87545}

\affiliation{$^3$National High Magnetic Field Laboratory,
Tallahassee, FL 32310}

\affiliation{$^4$Institute for Solid State Physics, University of
Tokyo, Kashiwa, Chiba 277-8581, Japan}

\affiliation{$^5$Theoretical Division, Los Alamos National
Laboratory, Los Alamos, NM 87545}

\begin{abstract}
We report an experimental determination of the phase boundary
approaching the quantum critical point separating a quantum
paramagnetic state and the proposed spin Bose-Einstein condensate of
triplons in the spin gap compound BaCuSi$_2$O$_6$. The ordering
temperature is related to the proximity to a quantum critical point
at the lower critical magnetic field $H_{\textrm{c1}}= 23.52 \pm
0.03 $ T by a power law parameterized by critical exponent $\nu$. We
obtain an experimental estimate of $\nu = 0.63 \pm 0.03$ down to a
temperature of 0.61K, which is in good agreement with the mean field
prediction of $\nu=\frac{2}{3}$ for the Bose-Einstein condensation
universality class.

\end{abstract}


\date{\today}

\maketitle

Several spin gap compounds, including BaCuSi$_2$O$_6$[1,2],
TlCuCl$_3$[3-7] and Sr$_2$Cu(BO$_3$)$_2$[8], have a singlet ground
state in zero magnetic field with a gap to the lowest excited
triplet state. The spin gap can be closed by an applied magnetic
field, such that a quantum critical point (QCP) at a magnetic field
$H_{\textrm{c1}}$ separates the quantum paramagnetic state from a
state characterized by long range magnetic order. The order
parameter for this transition is $\langle b^{\dagger}\rangle$, the
creation operator for a triplet state. In the absence of U(1)
symmetry-breaking anisotropy, the transition from the quantum
paramagnetic state ($\langle b^{\dagger}\rangle = 0$) to an ordered
state with broken U(1) symmetry ($\langle b^{\dagger} \rangle
\propto M^x_{\textrm{st}}+iM^y_{\textrm{st}}$ where
$M^x_{\textrm{st}}$ and $M^y_{\textrm{st}}$ are the $x$ and $y$
components of the staggered magnetization in the plane perpendicular
to the applied field) may be interpreted as a Bose Einstein
condensation (BEC) of triplons [9]. The magnetic field functions as
a chemical potential, thereby providing a convenient means to tune a
BEC to criticality at the QCP.

The spin gap system BaCuSi$_2$O$_6$ is unique because it provides
experimental access to a quantum phase transition of this nature in
a spin lattice that can be described by a U(1) rotationally
symmetric spin Hamiltonian [1]. A quantum phase transition that
belongs to the same universality class is observed in superfluid
$^4$He[10-12], but with a phase diagram that is comparatively
inaccessible for experimentation [13].

The proximity to the QCP is expected to be related to the ordering
temperature ($T_\textrm{c}$) by a power law $T_\textrm{c} \sim
(H-H_{\textrm{c1}})^\nu$ [14], which can be expressed in reduced
form
\begin{eqnarray}t &=&
f(h) \times (1-h)^\nu
\end{eqnarray}
where $t = \frac{T_{\textrm{c}}}{T_{\textrm{max}}}$, $h =
\frac{H_{\textrm{max}}-H}{H_{\textrm{max}}-H_{\textrm{c1}}}$
($H_{\textrm{max}}$ and $T_{\textrm{max}}$ represent the point on
the phase boundary halfway between $H_{\textrm{c1}}$ and
$H_{\textrm{c2}}$, the field at which the magnetization saturates)
and $f(h)_{h=1}$ is finite. The mean field critical exponent $\nu =
\frac{2}{3}$ is characteristic of the Bose-Einstein condensation
universality class, and describes the scaling behavior of a 3D
dilute interacting Bose gas near the QCP. The mean field estimate is
appropriate since the upper critical space dimension of 2
($d_\textrm{c} = 2$ because $z = 2$ for this universality class) is
exceeded [9, 15-17].

In this paper, we present a set of experiments on BaCuSi$_2$O$_6$
that examine the critical scaling of this phase transition in the
vicinity of the quantum critical point. The experimental results are
consistent with the $\nu = \frac{2}{3}$ BEC critical exponent, and
agree with Monte Carlo simulations in the lowest experimentally
accessible temperature window.

BaCuSi$_2$O$_6$ has a well characterized quasi 2D structure [1,2,18]
consisting of vertical Cu$^{2+}$ dimers arranged on a square
bilayered lattice staggered between bilayers. The inter-layer,
interdimer and interbilayer exchange couplings have been estimated
from high field magnetization data [1] to be $J = 4.45$meV, $J'
\simeq 0.58$meV, $J" \simeq0.116$meV respectively~[1,2].
Representing each dimer by a pseudospin with two possible states
(lowest energy states $s^z=0$ singlet and $s^z=1$ triplet), and
using appropriate transformations [1], the effective Hamiltonian
describes a 3D gas of hardcore bosons with in-plane nearest neighbor
hopping and repulsive interactions. At low temperatures, the system
undergoes a second order phase transition, which has been
interpreted as triplon condensation [1].

Here, magnetic torque, magnetocaloric effect and specific heat
measurements are performed to obtain points on the phase boundary
into the ordered state. Features in these thermodynamic quantities
characterise the classical 3D-XY phase transition into the ordered
state at finite temperatures.

Single crystal samples of BaCuSi$_2$O$_6$ grown by a flux-growth
technique [19] are used for these experiments, whereas previous
measurements on this material [1,2,18] used single crystals grown by
a floating zone technique. Flux grown crystals are chosen because of
a lower impurity content, a clearly defined Schottky anomaly in the
zero field heat capacity, and narrower nuclear magnetic resonance
lines. The characteristic features of the magnetic ordering
transition observed previously are identical in these samples,
indicating sample independence.

The specific heat of the sample is measured at $36$T in a $^4$He
cryostat in the hybrid magnet at Tallahassee. The inset to figure 1a
shows the characteristic lambda anomaly observed in specific heat,
indicating a second order phase transition into the ordered state.
The shape is identical to that in [1], which has been fit using
directed-loop Monte Carlo simulations. The ordering temperature at
$36$T is plotted on the phase diagram in figure 1a.

Magnetic torque measurements enable a sensitive probe of the
magnetic ordering transition at low temperatures. They are performed
in static magnetic fields up to $33$T in a $^3$He refrigerator in
Tallahassee. Samples are mounted on the moving plate of a phosphor
bronze capacitance cantilever, attached to a rigid plate rotatable
about an axis parallel to the axis of torque and perpendicular to
the applied magnetic field. The sample is mounted with a small angle
(< $10^o$) between the applied field and the normal to the sample
plane (easy axis $\hat{c}$), such that the applied field exerts a
torque on the crystal due to the difference in $g$-factor between
the $\hat{a}$ and $\hat{c}$ orientations with $g_a = 2.053 \pm
0.007$ and $g_c = 2.303 \pm 0.003$ [19]. The anisotropy in $g$
results in an anisotropy in $H_{\textrm{c1}} =
\frac{\Delta}{g\mu_B}$ (where $\Delta$ is the spin gap)[19]. Hence,
on entering the magnetically ordered phase with increasing magnetic
field, the anisotropy in magnetization causes a sudden increase in
torque, as the field attempts to align the $\hat{c}$ axis more
closely with the applied magnetic field. Torque measurements are
made during magnetic field sweeps across the ordering transition at
different temperatures.

The signature of the ordering transition is seen in field dependent
torque curves in a temperature range $0.61$K - $3.3$K (sample curves
shown in figure 1b). The field at which the phase transition occurs
is obtained from the position of a sharp feature in the second
derivative of the torque (example shown in the inset to figure 1b)
[20]. The feature becomes weaker at higher temperatures, but can be
extracted up to $T = 3.3$K. Examples of the ordering transitions
thus obtained are indicated by solid symbols on the torque curves in
figure 1b. Points on the phase diagram obtained from torque
measurements are shown as solid circles in the phase diagram plotted
in figure 1a.
\begin{figure}[htbp]
\includegraphics[width=0.5\textwidth]{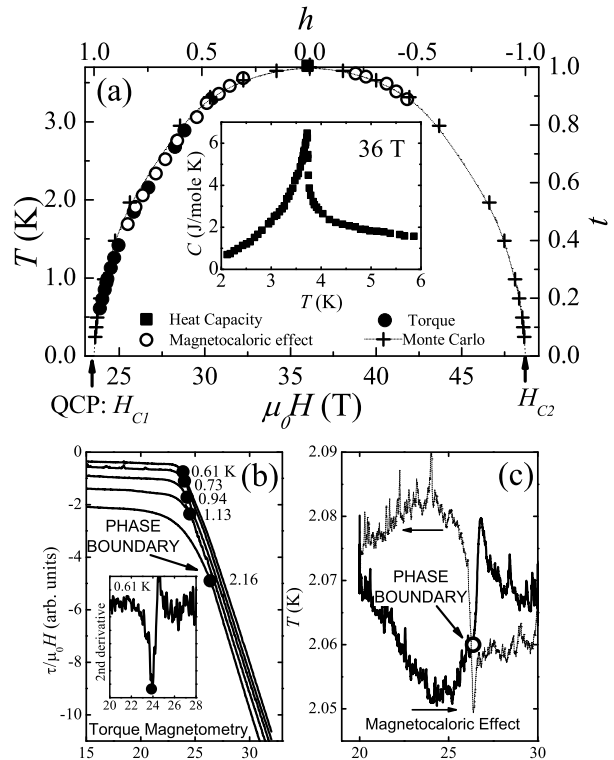}
\caption{(a) Points on the phase boundary are determined from
magnetic torque (solid circles), magnetocaloric effect (open
circles), specific heat (solid square) measurements and Monte Carlo
simulation ($+$ symbols, dotted line is a guide to the eye). The
inset shows the lambda anomaly in specific heat measured at a
magnetic field of 36T. (b) Torque measured as a function of rising
magnetic field at sample temperatures as indicated. The ordering
transitions determined from a sharp feature in the second derivative
(shown in the inset) are indicated on each of the torque curves. (c)
Sample curve indicating the temperature change due to the
magnetocaloric effect measured as a function of up and down sweeps
in a magnetic field. The ordering transition is indicated on the
curve.}
\end{figure}

The magnetocaloric effect describes the temperature change of a
magnetic material associated with an external magnetic field change
in an adiabatic process. An abrupt change in the temperature with
changing magnetic field indicates a large field variation of the
isothermal magnetic entropy, and is associated with an ordering
transition. These measurements are a good probe of ordering
transitions in a rapidly changing magnetic field, so can be used to
trace the phase boundary [21]. Magnetocaloric effect measurements
are carried out in fields up to $45$T in a $^4$He cryostat in the
hybrid magnet in Tallahassee. Temperature changes are detected
during magnetic field sweeps across the ordering transition at
different temperatures (shown in fig 1b). The peak in lattice
temperature at $H_{\textrm{c1}}$ ($H_{\textrm{c2}}$) in a rising
(falling) field indicates a drop in magnetic entropy with ordering.
Similarly, a dip in temperature in a falling (rising) field
indicates the transition out of the ordered phase at
$H_{\textrm{c1}}$ ($H_{\textrm{c2}}$). The position of the ordering
transition for $H < H_{\textrm{max}}$ is obtained from the onset of
the peak (as defined by the maximum in the first derivative) in a
rising field and for $H > H_{\textrm{max}}$ from the onset of the
dip in a falling field. The ordering transition thus obtained for a
representative field sweep is shown in figure 1b. Points on the
phase diagram obtained from magnetocaloric effect measurements are
shown by open symbols in the phase diagram plotted in figure 1a.

The Monte Carlo simulations for this system are performed using the
directed-loop algorithm [22] (results are represented by $+$ symbols
in figure 1a). Estimates of interdimer exchange coupling are refined
from [1], with revised values of $J' = 0.51$meV and $J" = 0.168$meV
yielding better agreement of the Monte Carlo simulations with
experimental points on the phase diagram.

\begin{figure}[htbp]
\includegraphics[width=0.51\textwidth]{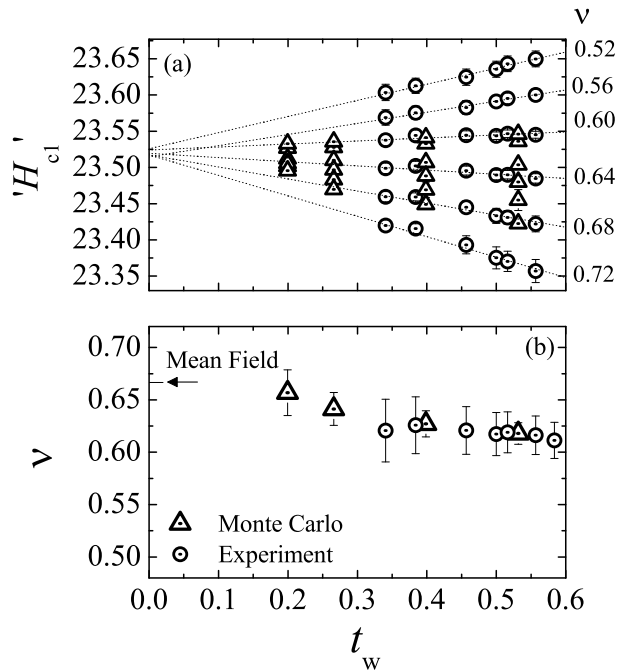}
\caption{(a) The circles represent estimates of $'H_{\textrm{c1}}'$
obtained from fitting the lowest few experimental points  on the
phase boundary in figure 1c in a window of increasing size
$t_{\textrm{w}}$, to eqn. 2 for different fixed values of $\nu$. The
x axis $t_{\textrm{w}}$ labels the highest reduced temperature of
the fit window. The dotted lines show the linear convergence of
$'H_{\textrm{c1}}'$ values at $t_{\textrm{w}} = 0$. The triangles
represent estimates of $'H_{\textrm{c1}}'$ similarly obtained from
Monte Carlo simulation data for corresponding fixed values of $\nu$,
and similar convergence is observed (b) The circles represent
estimates of $\nu$ from fitting the lowest few experimental points
on the phase boundary in figure 1a in a window of increasing size
$t_{\textrm{w}}$, to eqn. 2 with $H_{\textrm{c1}} = 23.52T$
determined from figure 2a. The error bars are due to experimental
uncertainty in determining values of $H_\textrm{c}$, leading to an
uncertainty in the value of $H_{\textrm{c1}}$. The triangles
represent estimates of $\nu$ from a similar fit to Monte Carlo
simulation data.}
\end{figure}

Experimental access to the particle-hole symmetric region in
BaCuSi$_2$O$_6$ enables us to extend the region near the QCP in
which the power law can be fit. Eqn. 1 describes scaling near
$H_{\textrm{c1}}$ while $t \propto (1 + h)^{\nu}$ describes scaling
near $H_{\textrm{c2}}$. In other words, the particle-hole symmetry
of the system implies that $t$ is a function of $h^2$:
\begin{eqnarray}
t = g(h^2) \times [(1 - h)(1+h)]^{\nu} &\equiv& g(h^2) \times (1 -
h^2)^{\nu}
\end{eqnarray}
where $g(h^2)$ varies more slowly than $f(h)$ in the vicinity of the
QCP.

In this set of experiments, points on the particle-hole symmetric
phase diagram first reported in [1] are refined using measurements
on a flux-grown single crystal and extended to low temperatures
close to the QCP (figure 1a). Critical scaling near the QCP is
described by eqn. 2. Importantly, we have access to sufficient data
points in the low temperature region near $H_{\textrm{c1}}$ to
obtain an estimate of the critical scaling exponent by fitting a
power law dependence in that region.

The power law dependence in the quantum critical region is extremely
sensitive to both the fit temperature range, and to the estimate of
$H_{\textrm{c1}}$ [17]. We use an empirical convergence approach to
determine the best estimate of $H_{\textrm{c1}}$. Figure 2a shows
the trend in the estimate of $H_{\textrm{c1}}$ (denoted as
$'H_{\textrm{c1}}'$) obtained by fitting the lowest few experimental
points  on the phase boundary in figure 1a in a window of increasing
size $t_{\textrm{w}}$ to eqn. 2 (where $g(h^2)$ is assumed to be
constant to a first approximation) for different fixed values of
$\nu$. Near the QCP, it is empirically observed from linear
extrapolation to $t_{\textrm{w}} = 0$, that estimates of
$'H_{\textrm{c1}}'$ become less dependent on $\nu$, and converge to
a single value irrespective of the value of $\nu$ (figure 2a).
Similar convergence to $H_{\textrm{c1}}$ is observed for the Monte
Carlo simulation results (figure 2a). The convergence is due to the
fact that the QCP is at $H_{\textrm{c1}}$, independent of the path
along which it is approached (characterized by $\nu$). From
$'H_{\textrm{c1}}'$ convergence, we obtain an estimate of
$H_{\textrm{c1}} \simeq 23.52 $T. The uncertainty in this value of
$H_{\textrm{c1}}$ is estimated to be $0.03$T, arising from the
experimental uncertainty in measuring $H_{c}$. The value of
$H_{\textrm{c1}} = 23.52 \pm 0.03$T thus obtained is then used to
estimate the critical exponent $\nu$.

The critical exponent $\nu$ is estimated from fitting eqn.2 (with
$g(h^2)\approx$~const.) to the narrowest temperature range near the
QCP with a statistically significant number of experimental
datapoints. Figure 2b shows the variation in $\nu$ with the size of
the temperature window that is fit to eqn. 2 (points on the phase
boundary are fit from the lowest value of $t = 0.16$ to a highest
value of $t = t_\textrm{w}$). As the critical region near the QCP is
approached with narrowing window size, $\nu$ approaches the
theoretical mean field value of $\frac{2}{3}$. In the lowest
experimentally accessible temperature window down to 0.61K, we
obtain a value of $\nu = 0.63 \pm 0.03.$ Performing a similar
analysis for data points from the Monte Carlo simulation reveals the
expected increase in $\nu$ to the theoretical mean field value as
the temperature window is further reduced below currently accessible
experimental temperatures (figure 2b). The experimental estimate of
$\nu = 0.63 \pm 0.03$ based on measurements down to temperature $t =
0.16$ is consistent with the theoretical mean field prediction of
$\nu = \frac{2}{3}$ to within experimental error.

\begin{figure}[htbp]
\includegraphics[width=0.51\textwidth]{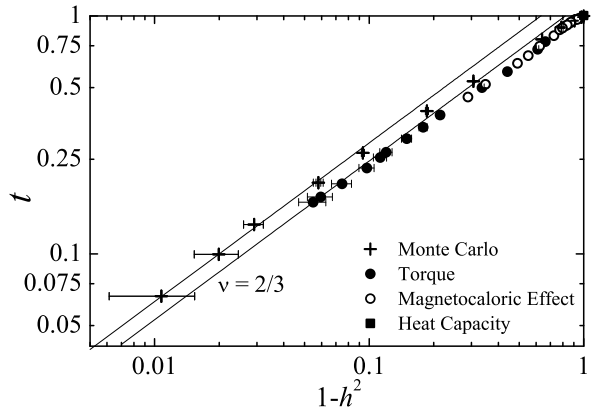}
\caption{Points on the phase boundary determined from magnetic
torque (solid circles), magnetocaloric effect (open circles) and
heat capacity (solid square) measurements for $H_{\textrm{c1}} =$
23.52 T, $H_{\textrm{max}} =$ 36.12 T, $T_{\textrm{max}}$ = 3.70 K
and from Monte Carlo simulation (crosses). The lines represent eqn.
2 with $\nu = \frac {2}{3}$.}
\end{figure}

Figure 3 shows the comparison between the Monte Carlo simulation and
the experimental data for $H_{\textrm{c1}} = 23.52$T. The lines
represent the power law (eqn. 2) with $\nu = \frac{2}{3}$. The Monte
Carlo simulation and experimental data are in good agreement in the
lowest experimentally accessible temperature window. We empirically
observe that the deviation from the power law at higher temperatures
is less in the case of the experimental data than for the Monte
Carlo simulation.

In summary, we performed magnetic torque and magnetocaloric effect
experiments to map out the phase diagram in the vicinity of the QCP
in the spin gap system BaCuSi$_2$O$_6$. Points down to 0.61K are fit
to the power law $t = g(h^2) \times (1 - h^2)^{\nu}$ (with
$g(h^2)\approx$~const.) to give a value $\nu = 0.63 \pm 0.03$. The
estimate we obtain for $\nu$ near the QCP is close to the
theoretical value $\nu$ = $\frac{2}{3}$, whereas previous
experimental measurements of $\nu$ in the spin gap system TlCuCl$_3$
have resulted in lower values in the range $0.43 - 0.60$ [3-7]. The
other known experimental measurements of the BEC critical exponent
$\nu$ have been on $^4$He adsorbed in aerogel [10-12], which is a
realisation of a dilute Bose gas, but is experimentally limited by
the presence of a random external potential. BaCuSi$_2$O$_6$ is a
unique U(1) symmetric spin gap material that enables experimental
access to a QCP separating a quantum paramagnet from a Bose-Einstein
condensate [1]. It provides a novel experimental realization of a
BEC in a grand canonical ensemble in the absence of an external
potential, with the region around the QCP accessible by a tuneable
external magnetic field.

This work is supported by the National Science Foundation (NSF),
DMR-0134613. Experiments performed at the NHMFL were supported by
the NSF, Florida State, and the Department of Energy. Monte Carlo
computations presented here were performed with the SGI 2800/384 at
the Supercomputer Center, Institute for Solid State Physics,
University of Tokyo. The numerical work was supported by a
Grant-in-Aid (Program No. 14540361) from Monkasho, Japan. S. E. S.
thanks D. I. Santiago for helpful discussions. I. R. F. acknowledges
support from the Alfred P. Sloan Foundation and S. E. S. from the
Mustard Seed Foundation.

\begin{acknowledgments}

\end{acknowledgments}


\begin{thebibliography}{1}
\bibitem{13}Tuning the particle hole density to provide access to the QCP
requires absorption of $^4$He in an external media, which introduces
a random external potential and effects due to surface absorption
that influence system behaviour.
\end{thebibliography}
\end{document}